\documentclass[preprint,prd,showpacs]{revtex4}
\usepackage{epsfig}
\begin{document}
\title{Kaluza-Klein Structure Associated With Fat Brane}
\author{Pham Quang Hung} 
\email[]{pqh@virginia.edu} 
\author{Ngoc-Khanh Tran} 
\email[]{nt6b@virginia.edu}
\affiliation{Department of Physics, University of Virginia \\
382 McCormick Road, Charlottesville, Virginia 22904-4714, USA}
\date{\today}
\begin{abstract}
It is known that the imposition of orbifold boundary conditions on background scalar field can give rise 
to a non-trivial vacuum expectation value (VEV) along extra dimensions, which in turn 
generates fat branes and associated unconventional Kaluza-Klein (KK) towers of fermions. 
We study the structure of these KK towers in the limit of one large 
extra dimension and show that normalizable (bound) states of massless and massive fermions can 
exist at both orbifold fixed points. Closer look however 
indicates that orbifold boundary conditions act to suppress at least half of bound KK modes, while 
periodic boundary conditions tend to drive high-lying modes to conventional structure. By investigating 
the scattering of fermions on branes, we analytically compute masses and wavefunctions of KK spectra 
in the presence of these boundary conditions up to one-loop level. Implication of KK-number 
non-conservation couplings on the Coulomb potential is also examined.
   \end{abstract}
\maketitle
   
\section{Introduction}
Conventionally, Kaluza-Klein (KK) towers of modes arise from the compactification of extra dimensions, 
i.e. the imposition of periodic boundary conditions on those finite extra dimensions, to render 
effectively the low-energy 4-d physics from a bigger spacetime with finite spatial extra dimensions 
(see e.g. \cite{Hamed2}). Given that Standard Model (SM) is the standard low-energy effective theory of 
particle interaction, one requires that any higher dimensional scenario reduces to the SM at the 
electroweak scale. From the KK perspective, this requires the lowest modes in KK towers of fermions 
to be chiral because these are the only ones we observe at low energy. To this end, the simple periodic 
compactification needs to be replaced by an orbifold compactification and, in the case of one extra 
dimension, the common choice is $  S^1/ Z_2 $ compactification. 

A novel mechanism has been proposed to localize Standard Model chiral fermions differently along 
extra dimensions making use of Yukawa interaction of these fields with a background scalar field 
\cite{Hamed1} (see also \cite{JR}). This approach is phenomenologically 
very attractive, because it allows for an easy control of the 4-d effective couplings by regulating the 
overlap of particles' wave functions in the extra dimensions. At the core of this mechanism, 
the necessary background scalar field of non-trivial VEV profile along the extra dimension has been 
explicitly realized through the imposition of various orbifold boundary conditions \cite{Georgi1} 
(see also \cite{KT}). Because of the nature of this mechanism, one sees that the KK 
modes obtained are intimately related to the structure of this scalar VEV in the transverse directions. 
At the same time, as in theories with finite extra dimensions, the usual periodic boundary conditions 
(i.e. periodic compactification) certainly drive KK towers back to the conventional structure. 
It is then of interest to analyze in details the interplay between these two tendencies revealed 
in the KK structure of related fields. 

For this purpose, we find it particularly useful to treat the dependence of fields on the extra 
dimensions as wavefunctions subject to the potential along the extra dimension 
associated with the background scalar kink solution (or ``bulk potential'' in short), 
whose centers are located at fixed points of the orbifold. In what follows, 
these defects will be referred to as fat branes.
In this view, we find that in the limit of a large extra dimension, bound (i.e. normalizable) KK 
states of massive fermions can actually exist at both fixed points, because 
bulk potential possesses local minima (or ``potential wells'') at those points (see Fig. \ref{Fig.2} below). 
This differs from the conclusion of \cite{Georgi1}, where in the same limit of a large extra 
dimension, normalizable KK states are found only at one of fixed points. This approach further 
allows us to make a direct connection between the boundary conditions imposed on wavefunctions and 
the transmission amplitude in a related scattering process of particles on the branes.

Our presentation is structured as follows.
Section II reviews the fermion localization mechanism \cite{Hamed1,Georgi1} in a model with one 
infinite extra dimension, where the extra dimensional wavefunction of fields and the corresponding KK 
mass can be exactly determined from the bulk potential viewpoint. In this limit, the periodic boundary 
conditions (PBCs) do not exist, so 
very distinct KK towers, partially discretized, originating from orbifold boundary conditions 
(OBCs) can readily be seen. 
Section III generalizes the calculation to the case of a large, but finite, extra dimension. 
Here the competition between the two types of boundary conditions becomes apparent. 
We show that, in this limit the OBCs essentially require wavefunctions along the extra dimension of 
any single 
KK mode to have the same parity at all fixed points, whereas bulk potentials tend to make them different at 
these points (see Fig. \ref{Fig.2}). As a result, this parity mismatch strongly affects KK light modes, 
while for the heavier KK modes, the same boundary conditions can be equivalently translated 
into the requirement that these KK fermions cross the branes without reflection. Hence in all cases, 
OBCs generally suppress many bound and low-lying KK modes. Section IV will highlight the fact that 
symmetry breaking effects in the transverse dimension are actually encoded in the effective couplings, 
which in turn generate non-universal and KK-number non-conservation interaction vertices of 
fermion-gauge boson and fermion-scalar. Sections IV-A and 
IV-B examine the implication of these effects on the Coulomb potential and the one-loop correction to KK 
masses respectively from a 4-d effective viewpoint. In a different but related problem and in order to 
clarify the physical meaning of the KK mass equation obtained 
in orbifold compactification, section V studies the scattering of fermion on branes. It is found that 
low-lying KK modes of fermion indeed survive only when the 
resonant transmision through the branes occurs, which in turn is achieved only for some very particular 
values of brane's parameters. Section VI summarizes our main results and offers some outlook. 
The appendix A classifies properties of general solutions of a hypergeometric-related differential equation 
encountered throughout the paper. 
    \section{Fermion localization in extra dimensions}
\begin{figure}
        \begin{center}    
        \epsfig{figure=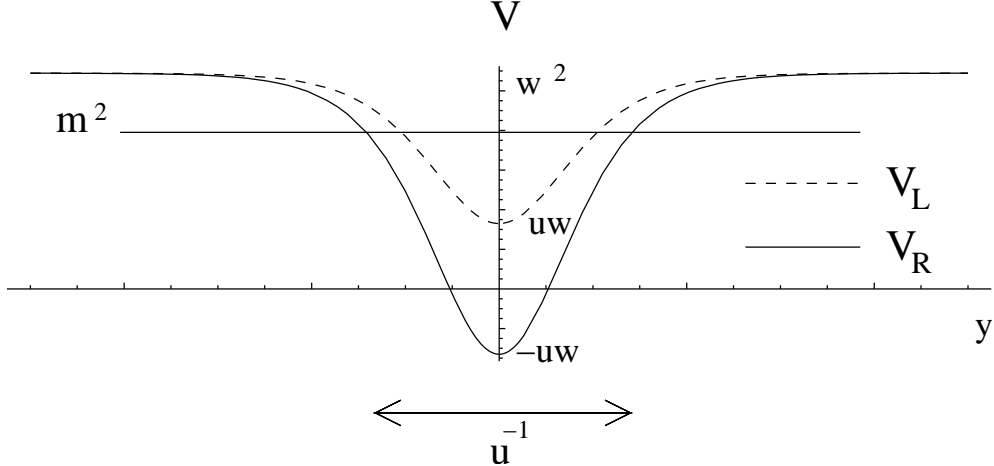,width=0.8\textwidth}
        \end{center}
        \vspace{0cm}
        \caption{Potentials $V_{R,L}$ experienced by fermions in the limit $L\rightarrow \infty $}
        \label{Fig.1}
        \end{figure}

Consider the case of a massless fermion and a real scalar field in 4+1 dimensions. We use $ y $ to 
denote the fifth dimension's coordinate, $y\in [0,L]$. In this section we will eventually let 
$ L\rightarrow \infty $.
The Lagrangians
         \begin{equation}
{\cal L}_{\psi} = \bar{\psi} (x,y) (i\gamma ^{\mu} \partial _{\mu} - \gamma ^{5} \partial _{y} - 
f\phi (x,y)) \psi (x,y) 
         \label{LagPsi}
         \end{equation} 
         \begin{equation}
{\cal L}_{\phi} = \frac{1}{2} \partial ^{\mu} \phi (x,y) \partial _{\mu} \phi (x,y) -  
\frac{1}{2} \partial _{y} \phi (x,y) \partial _{y} \phi (x,y) - \frac{\lambda}{4} (\phi ^2 (x,y) - v^2 )^2
         \label{LagPhi}
         \end{equation}
         \begin{equation}
\cal L = {\cal L}_{\psi} + {\cal L}_{\phi}
         \label{Lag}
         \end{equation}
are invariant under a $ Z_2 $ symmetry
         \begin{equation}
\phi (x,y) \rightarrow \Phi (x,y) \equiv - \phi (x,L-y)
         \label{Z2Phi}
         \end{equation} 
         \begin{equation}
\psi (x,y) \rightarrow \Psi (x,y) \equiv \gamma _{5} \psi (x,L-y)
         \label{Z2Psi}
         \end{equation}
which in turn allows the imposition of following orbifold and periodic boundary 
conditions
         \begin{equation}
\phi (x,-y) = \Phi (x,L-y) =  \phi (x,2L-y)
         \label{OBCPhi}
         \end{equation}
         \begin{equation}  
\psi (x,-y) = \Psi (x,L-y) =  \psi (x,2L-y)
         \label{OBCPsi}
         \end{equation}
We note that all fields are $2L$-periodic and the imposition of these boundary conditions actually 
transforms the extra dimension into an orbifold $ S^1 / Z_2 $ with two fixed points at 
$y=0$ and $y=L$. 
As $\phi (y)$ is antisymmetric at these points, if $\lambda v^2$ is sufficiently large, 
the minimization of effective potential
         \begin{equation} 
         \label{4dV} 
V(\phi)=\frac{1}{2} \partial _{y} \phi (x,y) 
\partial _{y} \phi (x,y) + \frac{\lambda}{4} (\phi ^2 (x,y) - v^2 )^2     
         \end{equation}
         \begin{equation}
         \label{VEVeq1}
\left. \frac {\delta V(\phi )}{\delta \phi }\right|_{h(y)} = - {\partial}^2 _{y} h + 
\lambda (h^2 - v^2) h = 0
         \end{equation}
gives rise to a non-constant solution $h(y)$ of VEV in the bulk. Furthermore, in the limit 
$L \rightarrow \infty$ considered in this section, it is just the well-known kink configuration
         \begin{equation}
h(y) = v \tanh { \sqrt{\frac{\lambda v^2}{2}}\,y}     
         \label{simplestsol}
         \end{equation} 
First substitute $\phi (x,y) = \langle \phi \rangle = h(y) $ in the leading approximation and then 
decompose $ \psi $ into 4-d chiral components $\psi _{R,L} (x)$ 
         \begin{equation}
\psi (x,y) = \psi _R (x) \xi _R (y) +  \psi _L (x) \xi _L (y)
         \label{chiraldecomp}
         \end{equation}
where the functions $\xi _{R,L} $ 
\footnote{The decomposition (\ref{chiraldecomp}) requires that $\xi _{R,L}(y) $ be continuous functions 
because they represent probability amplitude of finding fermions in extra dimension.}  
satisfy OBCs (\ref{OBCPsi})
         \begin{equation}
\xi _{R,L} (-y)   =  \pm \xi _{R,L} (y)    \;\;\;\;\;\;\;\;\;\;
\xi _{R,L} (L-y)  =  \pm \xi _{R,L} (L+y)    
         \label{OBCXi}
         \end{equation}
we obtain the equations of motion for 4-d chiral fields
         \begin{eqnarray}
i(\gamma ^{\mu} \partial _{\mu} \psi _L)\xi _L - \psi _R(\partial _y + fh)\xi _R & = & 0 \nonumber \\
i(\gamma ^{\mu} \partial _{\mu} \psi _R)\xi _R - \psi _L(-\partial _y + fh)\xi _L & = & 0
         \label{chiral5Diraceq}
         \end{eqnarray}   
The 4-d Dirac mass can be explicitly recovered when we take $ \xi _{R,L} \rightarrow \xi _{mR,L} $
with \cite{Georgi1,Hamed1}
         \begin{equation}
(\partial _y + fh)\xi _{mR}  =  m\xi _{mL}   \;\;\;\;\;\;\;\;\;\;
(-\partial _y + fh)\xi _{mL}  =  m\xi _{mR}
         \label{ansatz}
         \end{equation}
With $h(y)$ given in (\ref{simplestsol}), Eqs. (\ref{ansatz}) imply
         \begin{equation}
\left( -{\partial }^2 _{y} - uw \, \frac{1}{ \cosh ^2  {uy} } + w^2  \tanh ^2 {uy}  \right) \xi _{mR} 
= m^2 \xi _{mR}
         \label{QMXieqR}
         \end{equation}
         \begin{equation}
\left( -{\partial }^2 _{y} + uw \, \frac{1}{ \cosh ^2  {uy} } + w^2  \tanh ^2 {uy} \right) \xi _{mL} 
= m^2 \xi _{mL}
         \label{QMXieqL}
         \end{equation}
where we have defined
        \begin{equation}
w \equiv fv \; \; \; \; \; \; \; u \equiv \sqrt{\frac{\lambda v^2}{2}}
        \label{w}
        \end{equation}
In this ``Schr\"{o}dinger-like equation'', we see that $\xi _{mR,L}$ and the squared KK mass $m^2$ are 
respectively the eigenstates and eigenvalues subject to an ``analog bulk potential'' $V_{R,L} (y)$ 
generated by the background scalar VEV 
        \begin{equation}
        \label{chiralV}
V_{R,L} (y) \; = \; \mp  uw \, \frac{1}{ \cosh ^2 { uy}} + w^2 \tanh ^2 { uy }
        \end{equation}
The width of this ``potential well'' is $\sim \frac{1}{u} \sim \sqrt{\frac{1}{\lambda v^2} } $ and is 
the actual thickness of the domain wall (or fat brane) separating domains of different asymptotic 
values of scalar VEV along the fifth dimension. From Fig. \ref{Fig.1} we see that, if $u \geq w$, then $V_L$ 
becomes a ``potential hump'' and it possesses no bound states, thus for the sake of completeness, 
in the rest of this work we assume that $u$ is essentially smaller than $w$. 
Eqs. (\ref{QMXieqR}),  (\ref{QMXieqL}) are related to hypergeometric differential equation and can 
be exactly solved \cite{LL}. Technical details are given in Appendix A.

In the lower part of spectrum ($m^2 < w^2 $), bound KK masses are necessarily quantized
        \begin{equation}
        \label{mRquantized}
m^2 _n = 2n_R uw - n^2 _R u^2  \; \; \; \; \;  (0\leq n_R  < \frac{w}{u})  
        \end{equation}  
        \begin{equation}
        \label{mLquantized}
m^2 _n = 2(n_L +1) uw - (n_L +1)^2 u^2 \; \; \; \; \;  (0\leq n_L  < \frac{w}{u}-1)  
        \end{equation}  
where the ranges on $n_{R,L} \in N$ come from constraints $m^2 < w^2 $, $m _n \geq 0$.
The corresponding eigenfunctions can be written in term 
of hypergeometric functions and are given in (\ref{ARBsol}), (\ref{ALBsol}). The 
parity constraints (\ref{OBCXi}) on these wavefunctions further require $n_R$ and $n_L$ to be even and 
odd integer respectively. KK masses (\ref{mRquantized}), (\ref{mLquantized}) are already a surprising 
result, because in the case $L \rightarrow \infty$, the usual free KK spectrum 
which arises from compactification is continuous. Here, in contrast, KK spectrum is discretized 
for the first few levels ($n<\frac{w}{u}$) because the whole spectrum has been spontaneously distorted by 
non-trivial VEV through Yukawa interaction. In other words, the KK spectrum is now associated 
with the internal structure of fat brane, and not solely with the compactification.

In this bulk potential picture, for the zero mode $m_n = 0$, we have $n_R=0$ and no 
satisfying value of $n_L$, because (see Fig. \ref{Fig.1}) the $m^2 _n =0$ level 
is even lower than the lower limit 
of potential $V_L$, so only right-handed zero mode is normalizable and survives. 
\footnote{By flipping the sign of 
$\gamma _5$ in (\ref{Z2Psi}), one can inverse the situation where now only left-handed 
zero mode survives.} 
This is one of the distinct features of this model, i.e. it possesses only single-chirality zero-mode 
fermions, which then can be identified with the Standard Model chiral fermions. Various interesting 
phenomenological applications of fat brane models, such as fermion mass hierarchy, 
CP violation, baryogenesis and proton decay suppression, etc. have been carried out in numerous works 
(see e.g. \cite{Hamed1}, \cite{KT}, \cite{FBapplication}). 

The non-zero KK modes of mass $m_n$ come in pair of both chiralities with the relation 
(\ref{mRquantized}), (\ref{mLquantized})
         \begin{equation} 
         \label{nRL}
          n_R = n_L + 1
         \end{equation}
Obviously, this assures the difference in parity of $\xi _{R,L}$ as earlier required by OBCs. In the case 
$L$ is finite, relation (\ref{nRL}) is rather strict that it apparently acts to suppress many light KK modes 
as we will see in the next section.

We now proceed to the continuous spectrum, $m^2 \geq w^2 $, $k \equiv \sqrt{\frac{m^2 - w^2}{u^2}}$. 
With the parity constraints (\ref{OBCXi}) we can unambiguously solve for corresponding wavefunctions 
(see (\ref{AReven}), (\ref{ALodd})), whose asymptotic forms are 
         \begin{eqnarray}
         \label{assymR}   
y\rightarrow \pm \infty & : &  
\xi _{mR} (y) \rightarrow 2^{-ik} (e^{\pm ikuy} + \frac{D_1}{1-D_2}e^{\mp ikuy} )     \\
         \label{assymL}
y\rightarrow \pm \infty & : &  
\xi _{mL} (y) \rightarrow \pm 2^{-ik} (e^{\pm ikuy} - \frac{D'_1}{1+D'_2}e^{\mp ikuy} )
         \end{eqnarray}
where $D_1$, $D_2$, $D'_1$, $D'_2$ are given in (\ref{ACD}), (\ref{AC'D'}). 
Clearly, under the refection $y\rightarrow -y $, $\xi _{mL}$ changes its signs but $\xi _{mR}$ does not, 
i.e. they are manifestly antisymmetric and symmetric respectively at $y=0$ in these forms. However, 
we see also that matter waves propagate in both directions even as $y \rightarrow \pm \infty$ 
signaling the existence of other fixed points at infinity, from which waves reflect. But in the 
limit $L \rightarrow \infty$ considered here, this reflection does not seem to be very reasonable. 
Either ways, this is not more problematic, because the model of infinite extra dimension itself is not 
realistic as soon as we introduce gauge fields that can propagate in the bulk. For now we just note that
the reflection-from-infinity is a consequence of orbifold boundary condition (\ref{OBCPsi}) imposed on 
fermion fields. In the next sections we investigate the case of finite extra dimension and this problem 
will no longer exist.

Similarly, one can analyze the KK expansion of background scalar field $\phi (x,y)$ about its VEV $h(y)$
(\cite{Georgi1})
        \begin{equation}
        \label{scalarexpansion}  
\phi (x,y) = h(y) + \sum_{n} \phi _n (x) f_n (y)        
        \end{equation}     
The orbifold boundary condition (\ref{OBCPhi}) forces $f_n (y)$ to be antisymmetric at all fixed points.
        \begin{equation}
        \label{OBCf}
f_n (y)=-f_n (-y)  \; \; \; \; \; \; \;  f_n (L+y)=-f_n (L-y)
        \end{equation}
Plugging the expansion (\ref{scalarexpansion}) into scalar action and keeping only terms up to quadratic 
order in $\phi _n$, we obtain
        \begin{equation}
        \label{scalaraction}
S= \int d^4 x \sum_{m,n} \left( \frac{1}{2} \partial ^{\mu} \phi _m \partial _{\mu} \phi _n
\int dy f_m(y) f_n (y) - \frac{1}{2}\phi _m \phi_n \int f_m(y) (-\partial ^2 _y + \Delta^2 (y))f_n(y) 
\right)
        \end{equation} 
where
        \begin{equation} 
        \label{Delta}
\Delta ^2 (y)=\left. \frac{\partial ^2}{\partial \phi^2} \right|_{h(y)} 
\left( \frac{\lambda}{4}(\phi ^2 -v^2)^2 \right) 
= \lambda (3h^2 -v^2)
        \end{equation}
In the case $L\rightarrow \infty$ we have (\ref{simplestsol})
        \begin{equation} 
        \label{inftyDelta}
\Delta ^2 (y)=\lambda v^2 (3 \tanh ^2 {uy} -1)
        \end{equation}
As before, the 4-d Klein-Gordon mass of KK scalar modes is recovered when $f_n$ satisfies
        \begin{equation}
        \label{QMfeq}
\left(-\partial^2_y+\Delta^2(y)\right)f_n(y)=\bar m^2_n f_n(y)
        \end{equation}
Evidently, solutions of Eq. (\ref{QMfeq}) are orthogonal to one another, and after integrating over $dy$ 
we are left with a tower of 4-d scalars $\phi_n$ of mass $\bar m _n$ in (\ref{scalaraction}). 
The general solutions of (\ref{QMfeq}) are given in Appendix A.

For the discrete spectrum, $\bar \varepsilon \equiv \sqrt{\frac{4u^2 - \bar m ^2}{u^2}}$, we have only 
one bound eigenstate antisymmetric at $y=0$ (\ref{Afsol}) of 
quantum numbers (\ref{Anf}) $\bar n =1$, $\bar \varepsilon =1$, 
        \begin{equation} 
        \label{discretsolf}
f(y)=\frac{1}{\cosh{uy}} F(-1,4,2;\frac{1}{1+e^{2uy}})=\frac{\tanh{uy}}{\cosh{uy}}
        \end{equation}
corresponding to KK mass $\bar m =\sqrt {3} \, u$. 

For the continuous spectrum, $\bar m ^2 \geq 4u^2$, 
$\bar k \equiv \sqrt{\frac{\bar m ^2 - 4u^2}{u^2}}=i\bar{\varepsilon}$, the 
wavefunction $f_n(y)$ being antisymmetric at $y=0$ can be found (see(\ref{Afodd})). Its asymptotic 
form is
        \begin{equation} 
y\rightarrow \pm \infty  : \;\;\;\;\;\;\;  
f_n (y) \rightarrow \pm 2^{-i\bar k} \left( e^{\pm i\bar k uy}
- \frac{\Gamma (-i\bar k) \Gamma (1-i\bar k)}{\Gamma (-i\bar k -2) \Gamma (-i\bar k +3)}
e^{\mp i\bar k wy} \right)
        \label{assymf}
        \end{equation}  
where $\bar D$ is given in (\ref{ADbar}).
This shows that, just like the case of fermion mentioned below Eq. (\ref{assymL}), the orbifold 
boundary condition (\ref{OBCf}) gives rise to the non-physical reflection-from-infinity of background 
scalar wavefunctions in the extra dimension. This point will be clarified below in a more 
realistic scenario where the other orbifold fixed point is taken into account by considering large, 
but not infinite $L$.  
\section{Finite limit of extra dimension }
In the model of infinite and flat extra dimension investigated in the last section, KK 
masses and wavefunctions can be solved exactly, but it may not be realistic as soon as the SM
gauge bosons or the graviton are introduced
\footnote{ Theories with infinite but warped extra dimensions effectively consistent with 4-d 
physics have been built, see \cite{RS}.}. In this section we 
consider a more realistic situation where $L$ is no longer infinite (see also \cite{Georgi1}). 
It turns out that in a certain 
limit of large but finite $L$, we are still able to carry out the computation with a 
very good approximation. 
First we note that all potentials of the forms (\ref{chiralV}), (\ref{inftyDelta}) have the same 
characteristic width $\sim \frac{1}{u}=\sqrt{\frac{2}{\lambda v^2}}$, so if  
$L$ is several times larger than $\frac{1}{u}$, the two potential wells centered at $y=0$ and
$y=L$ are neatly separated. Consequently, we can solve the eigenstates of this double-well potential 
by joining eigenstates of individual wells together. Indeed when 
$L\gg \sqrt{\frac{2}{\lambda v^2}}$, the corresponding solution of (\ref{VEVeq1}) antisymmetric at $y=0$, 
$y=L$ (by OBCs (\ref{OBCPhi})) is
        \begin{equation}        
        \label{compositesol}
h(y) = v \tanh {\left(\sqrt{\frac{\lambda v^2}{2}}\,y \right )} 
\tanh {\left (  \sqrt{\frac{\lambda v^2}{2}}\,(L-y) \right )} + 
{\cal{O}} (e^{-L{\sqrt{\lambda v^2}}})
        \end{equation} 
which together with (\ref{ansatz}) gives rise to new composite potentials   
        \begin{equation}
        \label{inftychiralV}
V_{R,L} (y) = \mp  uw  \left( \frac{\tanh{u(L-y)}}{ \cosh ^2  {uy} } - \frac{\tanh {uy}}
{ \cosh ^2  {u(L-y)} }\right) + w^2  \tanh ^2 {uy}  \tanh ^2 {u(L-y)}
        \end{equation} 
        \begin{figure}
        \begin{center}
        \epsfig{figure=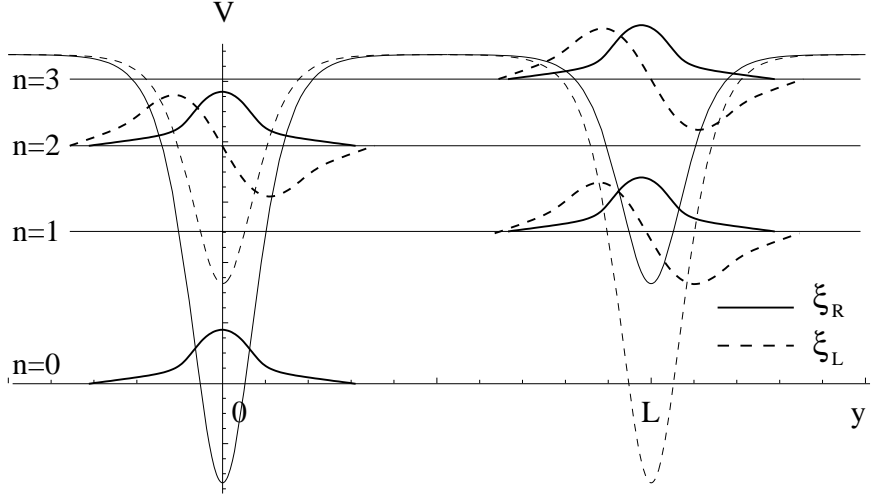,width=0.7\textwidth}
        \end{center}
        \vspace{0cm}
        \caption{Potentials $V_{R,L}$ and symbolic wavefunctions $\xi _{R,L}$ of fermion chiral 
         components with $L=\frac{10}{u}$}
        \label{Fig.2}
        \end{figure}
In Fig. \ref{Fig.2} we sketch the potentials (\ref{inftychiralV}) for the case 
$L=10 \sqrt{\frac{2}{\lambda v^2}}$ (to be compared with Fig. \ref{Fig.1}). 
The important observation here is that, 
the potentials $V _R$ and $V _L$ exchange their shapes at fixed points, where only one of two first terms in 
(\ref{inftychiralV}) contributes significantly. We will now attempt to compute the KK masses 
and corresponding wavefunctions in this limit through a matching process. 
Eqs. (\ref{QMXieqR}), (\ref{QMXieqL}) become
        \begin{equation}    
        \label{inftyQMXieq}
\left( -{\partial }^2 _{y} + V_{R,L}(y) \right) \xi _{mR,L} = m^2 \xi _{mR,L}
        \end{equation}
where $V_{R,L}$ now are given in (\ref{inftychiralV}).
For the light modes, $m^2 < w^2 $, $ \varepsilon \equiv \sqrt{\frac{w^2-m^2}{u^2}} >0 $, 
wavefunctions which are bound to each component well are given in (\ref{ARBsol}), 
(\ref{ALBsol}), all of which decay to zero as fast as $\exp{(-\frac{\varepsilon uL}{2})}$ at the region 
separating the two wells ($y \sim \frac{L}{2}$). So as far as the amplitude is 
concerned, the matching of two individual wavefunctions of the same $m^2$ is automatic. 
However, the parity matching is not so obvious. At the same squared mass level $m^2$, $\xi _{mR}$ 
behaves like component right-handed wavefunction at $y=0$ and left-handed wavefunction at $y=L$, so 
in view of (\ref{OBCXi}), if $\xi _{mR}$ is symmetric at $y=0$, it must be antisymmetric at $y=L$ or 
vice-versa. Whereas OBCs (\ref{OBCXi}) also require that $\xi _{mR}$ be symmetric 
at all fixed points. So if $\xi _{mR}$ were a solution satisfying these apparently contradictory 
conditions on parity and were non-zero in the vicinity of $y=0$, then it should be zero around $y=L$, 
because null function is the only function being both symmetric and antisymmetric with respect to a 
given reference point. And the same parity mismatch holds for left-handed states. As $\xi _R$, $\xi_L$ are 
respectively symmetric and antisymmetric at fixed points, Fig. \ref{Fig.2} indicates that the tower of 
bound (right and left-handed) KK states with mode's indices $n=0,2,...$ will be localized 
at $y=0$, while tower with $n=1,3,...$ localized at $y=L$. 
\footnote{In \cite{Georgi1}, after combining (\ref{ansatz}) and 
(\ref{compositesol}) to get the zero-mode wavefunction 
$ \xi _{0R} = {\cal {C}} (\cosh {u(L-y)})^{\frac{fv}{u}} \sim e^{fv(L-y)}$ close to fixed point $y=L$, it 
is accordingly found that the zero and other massive KK modes are non-normalizable at $y=L$. 
Though $\xi _{0R}$ increases exponentially as y runs away from $y=L$, we think that the matching 
of this function with zero-mode state (\ref{ARBsol}) bound to fixed point $y=0$, which decreases 
exponentially into the bulk, would necessarily set the constant ${ \cal {C}}$ to zero in order to preserve 
the normalization of $\xi _{0R}$ in the limit of large L. As a result, the right-handed zero mode exists  
only around $y=0$ while certain higher normalizable modes can exist around both fixed points as seen 
in the bulk potential picture (Fig. \ref{Fig.2}).   } 
These towers do not essentially correlate, but in the limit $L \sim \frac{1}{u} $ stronger tunneling may 
change the qualitative picture. 

We can see this more clearly now by studying the upper part of spectrum, where 
$m^2 \geq w^2 $, $k \equiv \sqrt{\frac{m^2-w^2}{u^2}}=i\varepsilon$. Here the matching process is simple, 
because as noted earlier, at the matching region in between the two wells, the wavefunctions associated 
with each well already fully reach their asymptotic forms.
    
Let us first consider the right-handed modes (symmetric at all fixed points). When the fixed points at 
$y=0$ and $y=L$ are respectively taken to be the origin of the y-coordinate, the symmetric 
wavefunction $\xi _R$ at the matching point 
($y \sim \frac{L}{2}$) reads (see (\ref{AReven}), (\ref{ALeven}))
        \begin{equation}
\xi _{R} (y) \sim  e^{ikuy} + \frac{D_1}{1-D_2}e^{-ikuy} 
        \label{MassymR}
        \end{equation}
        \begin{equation}
\xi _{R} (-L+y)=\xi _{R} (L-y) \sim  e^{iku(L-y)} + \frac{D'_1}{1-D'_2}e^{-iku(L-y)} 
        \label{MassymR'}
        \end{equation}    
The matching of (\ref{MassymR}) and (\ref{MassymR'}) induces the following relation
        \begin{equation}
\frac{D_1}{1-D_2} =  e^{2ikuL}  \frac{1-D'_2}{D'_1} 
        \label{relation1}
        \end{equation}        
Since $\xi _{R}$ is 2L-periodic (\ref{OBCPsi}), we can rewrite (\ref{relation1}) as
        \begin{eqnarray}
        \label{Mcondition1}
e^{2ikuL} = 1 \Leftrightarrow ku &=& \frac{n \pi}{L} \\
        \label{Mcondition2}
D_1 D'_1+{D_2}^2 &=&  1
        \end{eqnarray}
The first of these equations is the usual periodic boundary conditions, while the second accounts 
for orbifold boundary conditions and effect of fat branes. Because $k$, $D'$s all are functions of $m$, 
(\ref{Mcondition1}), (\ref{Mcondition2}) are actually equations determining KK masses. 

Matching antisymmetric left-handed wavefunctions produces relations identical to (\ref{Mcondition1}) and 
(\ref{Mcondition2}), this ensures the equality of 4-d left and right-handed fermion masses in the 
same KK mode. Remarkably, the physical meaning of condition (\ref{Mcondition2}) can be viewed as the 
requirement that the transmission amplitude of fermions in the double-brane system needs to 
be exactly one. 
Quantitative discussion will be given in section V. However the assertion's outcome itself is readily 
plausible: once the transmission amplitude is one, the KK states propagate in the bulk as if there were 
no potential at all, and subsequently the periodic boundary condition (\ref{Mcondition1}) drives them to 
the conventional structure. As a result, we expect high-lying modes of KK spectrum to have structure 
closer to that of periodic compactification.

Let us now examine the compatibility of two conditions (\ref{Mcondition1}), (\ref{Mcondition2}). 
For very heavy KK modes ($k \gg \frac{w}{u} >1$), $D_1$, $D'_1$ and $D_2$, $D'_2$ approach one 
and zero respectively, and the condition (\ref{Mcondition2}) is automatically satisfied. Then 
(\ref{Mcondition1}) implies 
$ m^2=w^2+k^2 u^2= w^2 + \frac{n^2 \pi ^2}{L^2} \approx \frac{n^2 \pi ^2}{L^2} $, which is the usual 
KK mass from periodic compactification. For smaller values of $k$, it may be difficult 
to solve (\ref{Mcondition2})
analytically although a numerical approach is possible. In section V we will discuss the solution of this 
equation in the limit of small $k$. Here we just mention that for $k$ which is not very large, the 
two conditions 
(\ref{Mcondition1}), (\ref{Mcondition2}) are not always compatible. To illustrate this point, let us 
consider a special case where $s \equiv \frac{w}{u}$ is an integer. Using 
$\Gamma (z+1)=z \Gamma (z)$ and (\ref{ADrelations}), we find $D_2=D'_2=0$ and 
         \begin{equation}
         \label{sIntegral}
D_1=\frac{(ik+1)\ldots(ik+s)}{(ik-1)\ldots(ik-s)} \; \; \; \; \; \; \; 
D'_1=\frac{(ik+1)\ldots(ik+s-1)}{(ik-1)\ldots(ik-s+1)}
         \end{equation}
Eq. (\ref{Mcondition2}) becomes a polynomial equation of order $(k^2)^{s-1}$, so it has at most $(s-1)$
solutions of $k^2$ (or squared mass $m^2$), which may or may not satisfy (\ref{Mcondition1}). 
In any cases, the combination of (\ref{Mcondition1}), (\ref{Mcondition2}) gives no more than $(s-1)$ 
different exact values of $m^2$ in the range of not-too-large $k$ ($k\leq s \equiv \frac{w}{u} $), while in 
a conventional compactification (\ref{Mcondition1}) the number of KK states in the same range is 
$\sim \frac{wL}{\pi} \gg 1$. 
\footnote{However, we will see later that the case of integral $w/u$ is a special ``resonant'' case, 
where Eq. (\ref{Mcondition2}) is approximately satisfied for very small values of k.}
Combined with the earlier consideration for bound KK states, we see 
clearly that the lower part of KK spectrum is strongly distorted by spontaneous breaking of 
background scalar, to which fermions are coupled, and also by orbifold boundary 
conditions, and the latter are generally strict that many of these states are effectively suppressed. 
The higher levels are not essentially affected by 
bulk potential (their transmission coefficient is close to unity) and usual periodic 
boundary conditions drive their mass structure to that of conventional compactification.

\begin{figure}
        \begin{center}
        \epsfig{figure=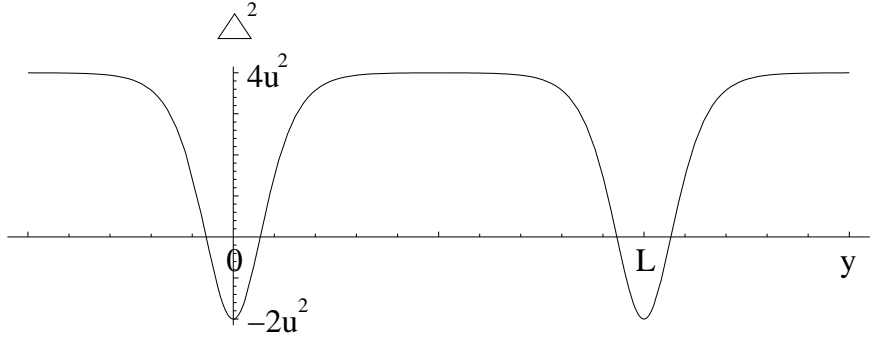,width=0.7\textwidth}
        \end{center}
        \vspace{0cm}
        \caption{Potentials $\Delta ^2(y)$ experienced by scalar fields with $L=\frac{10}{u}$}
        \label{Fig.3}
        \end{figure}

We now briefly investigate the KK scalar structure (\ref{scalarexpansion}), (\ref{QMfeq}) in the limit 
$L\gg \frac{1}{u}$ by a similar method. In this limit, the potential (\ref{inftyDelta}) becomes 
(see(\ref{compositesol}))
         \begin{equation} 
         \label{compositeDelta}
\Delta ^2 (y)=\lambda v^2 (3 \tanh ^2 {uy} \tanh ^2 {u(L-y)} -1 )
         \end{equation}
This potential is sketched in Fig. \ref{Fig.3} for $L=\frac{10}{u} = 10 \sqrt{\frac{2}{\lambda v^2}} $. 
We see that two component wells are identical (i.e. the potential is L-periodic) and no parity mismatch 
occurs. Further, in the limit $L\gg \frac{1}{u}$, discrete scalar KK levels bound to the first well are 
not interfered by those bound to the second and still we have only one antisymmetric KK state 
(\ref{discretsolf}) with mass $\bar m={\sqrt 3}u$ in the lowest part of spectrum. 
For the higher modes, ${\bar{m}}^2 \geq 4u^2$, the matching of scalar wavefunctions $f(y)$ (\ref{Afodd}) 
antisymmetric at all fixed points generates following relations
        \begin{eqnarray}
        \label{Mcondition1S}
\bar k u &=& \frac{\bar n \pi}{L} \\
        \label{Mcondition2S}  
{\bar D}^2 &=& \left ( \frac{(1+i\bar k)(2+i\bar k)}{(1-i\bar k)(2-i\bar k)} \right ) ^2 =1  
        \end{eqnarray}   
where $\bar{k} \equiv \sqrt{\frac{\bar{m}^2-4u^2}{u^2}}$.
These equations have at most two exact solutions $\bar k ^2 =0$ and $\bar k ^2 = 2 $ (the latter is a 
true solution if only $\frac{\sqrt 2 Lu}{\pi}$ is integer) corresponding to ${\bar m}^2=4u^2 $ and 
$6u^2$. However, for all $\bar k \ll 1$ or $\bar k \gg 1$, the condition (\ref{Mcondition2S}) is 
approximately satisfied and in these ranges, only the periodic boundary condition 
(\ref{Mcondition1S}) effectively governs the KK masses and gives them the conventional structure up to 
a constant shift
        \begin{equation} 
        \label{Smass2ranges}
{\bar m _n}^2=u^2(4+{\bar{k}}^2)=u^2(4+ \frac{\bar{n} ^2 \pi ^2}{u^2L^2})
        \end{equation}
where $\bar n \gg uL/ \pi$ or $\bar n \ll uL/ \pi$. As scalars are not chiral, they do not 
suffer from parity mismatch and consequently their transmission coefficient can reach to unity not 
only for high-lying KK levels, but also for levels immediately above the surface of potential well. 
This ``resonant behavior'' of potential (\ref{inftyDelta}) was discovered long time ago 
(see \cite{Voloshin}). We 
note that, in the difference with the result obtained therein, here scalar fields are constrained to 
be antisymmetric by OBCs, then resonance occurs only in two above ranges of momentum $\bar k$. This 
special structure of KK scalar spectrum has important implications in practical calculation as 
we will see in the following fermion self-energy evaluation.  

\section{Four-dimensional effective couplings and implications}
It is well known that in universal extra dimension (UED) scenario (see e.g. \cite{UED}) where no 
localization mechanism is invoked and all extra dimensions are accessible to all fields, momentum is 
conserved in both longitudinal (infinite) and transverse (finite) directions of space. This in turn 
implies the KK-number conservation of all vertex interactions and there are no tree-level contributions 
from KK excitation to Standard Model observables, whose content fields are taken to be the zero-modes in KK 
tower picture. The situation is quite different in brane scenario, where Lorentz invariance is violated 
along transverse dimensions due to both background kink and orbifold 
compactification. Consequently, in the reduced 4-d picture, there exist KK-number non-conservation vertices 
characterized by effective couplings. This is because the overlap-integration 
over extra coordinates leading to 4-d couplings actually measures the effects 
of Lorentz invariance breaking left on the wavefunctions of related fields along the extra dimension. These 
couplings may give rise to new interesting phenomenologies such as tree-level flavor changing neutral 
current, new mixing of quark and lepton flavors, etc. The comparison of these new contributions with 
experimental data will provide bounds on various parameters of the model. In this section, just for the 
purpose of illustration, we will present a toy model which involves only one fermion flavor in order 
to study the new implications, if any, of brane scenario to Coulomb interaction in a tree-level 
consideration and to the KK masses up to one-loop level. 
\subsection{Static Coulomb Interaction}    
\begin{figure}
        \begin{center}
        \epsfig{figure=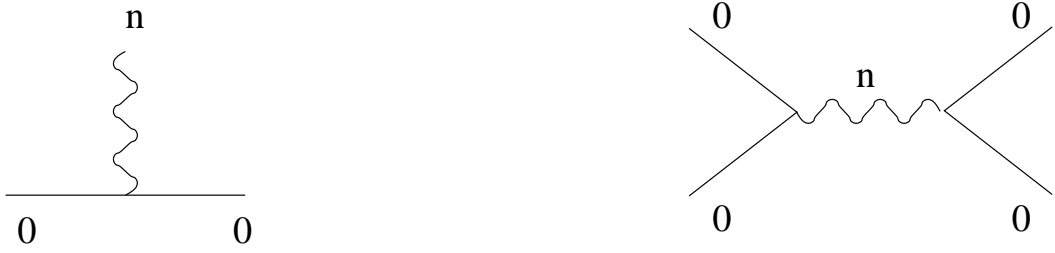,width=0.85\textwidth}
        \end{center}
        \vspace{0cm}
        \caption{Effective fermion-photon vertex and tree-level fermion scattering diagram}
        \label{Fig.4}
        \end{figure}

Charged fermions in QED interact with one another by exchanging a photon. We now assume that the 
photon can propagate freely along the fifth dimension, 
otherwise the localization mechanism for gauge boson in the exact fashion as we did for fermions 
encounters a serious complication related to the issue of universality of charge. 
The $ S^1 / Z_2 $ OBCs turn the 
extra dimensional wavefunction of the photon KK modes in the form of Cosine or Sine functions depending 
on their transformation property under $Z_2$. The usual choice of photon's first four components 
$A_{\mu}$ being symmetric and the fifth $A_5$ antisymmetric at $y=0,L$ eliminates the zero mode of 
$A_5$ as well as the contribution of its KK excitations $A_{n5}$ to zero-mode charged fermions' 
interaction (Fig. \ref{Fig.4}).
After dimensional reduction, we obtain the following vertex coupling of fermion zero-mode and photon 
KK modes:
            \begin{equation}
            \label{LagQEDvertex}
\int{dy \bar{\psi}(x,y) (-e_{5}\not{A}(x,y)) \psi (x,y)} \rightarrow 
-e\bar{\psi}_{0}(x) \gamma^{\mu} A_{0\mu}(x) \psi_{0}(x)
-\sum_{n=1}^{\infty}\epsilon_{n}\bar{\psi}_{0}(x) \gamma^{\mu} A_{n\mu}(x) \psi_{0}(x)
            \end{equation} 
where $\epsilon_{n}$'s are 4-d effective couplings            
            \begin{equation}
            \label{charge4-d}
\epsilon_{n} \equiv \frac{e_{5}}{\sqrt{L}}\int_{0}^{2L} dy (\xi_{0}(y))^{2} \cos{(\frac{n\pi y}{L})}
            \end{equation}
and $e={e_{5}}/{\sqrt{2L}}$ is the usual 4-d charge of fermion zero mode, whose wave function is 
$\xi_{0}(y)=(\cosh{uy})^{-w/u}$ as given in (\ref{ARBsol}). In the non-relativistic limit, the potential 
between two charged fermion zero modes can be found by working out the KK photon 
exchange process depicted in Fig. \ref{Fig.4}. The result is
            \begin{eqnarray}
            \label{CoulombV}
V(r)&=&\frac{e^2}{r}+\sum_{n=1}^{\infty}\frac{\epsilon_{n}^2}{r}e^{-n{\pi}r/L} \\ \nonumber
&=&\frac{e^2}{r}+\frac{2e^2}{r}\int_{0}^{2L}dy(\xi_{0}(y))^2\int_{0}^{2L}dy'(\xi_{0}(y'))^2
\sum_{n=1}^{\infty}\cos{(\frac{n\pi y}{L})}\cos{(\frac{n\pi y'}{L})}e^{-n{\pi}r/L}
            \end{eqnarray}
where $r$ denotes the spatial separation in 4-d picture. The first term of (\ref{CoulombV}) presents 
the contribution of massless photon, while remaining terms come from its massive KK modes. By 
transforming the sum over mode's index $n$ into the sum over elements of a geometric series, and 
approximating $\xi_0(y)$ to a Gaussian function (\ref{Gaussianfn}), we obtain the following 
expression for the potential $V(r)$:
            \begin{equation}
            \label{leadingV}
V(r)=\frac{e^2}{r}+\frac{2e^2}{r}\left(-1+\frac{1}{1-e^{-{\pi}r/L}}-\frac{3{\pi}^2}{4wuL^2}
\frac{e^{-{\pi}r/L}(1+e^{-{\pi}r/L})}{(1-e^{-{\pi}r/L})^3} + 
{\cal O}(\frac{1}{w^2u^2L^4})(\frac{L}{r})^5 \right)
            \end{equation}  
If $r\gg L$, (and $r > 1/M_Z$ so that the contribution from the Z-boson can be neglected), 
the potential (\ref{leadingV}) takes the form 
            \begin{equation}
            \label{V(r>L)}
V(r)\simeq \frac{e^{2}}{r}+(1-\frac{3{\pi}^2}{4wuL^2})\frac{2e^2}{r}e^{-{\pi}r/L} + \ldots
            \end{equation}
where we have kept only a few terms of higher orders. The first term in (\ref{V(r>L)}) is the usual 
Coulomb interaction potential arising from the exchange of the zero mode 
of the photon, whereas the second term has a Yukawa potential form because it is mediated by an 
infinite tower of massive KK modes. This correction is considerable only when $r$ approaches 
the size of the extra dimension, and the brane's explicit contribution to this 
correction is at most few percents in the limit $L \gg u$, so it is unlikely that we could find some 
phenomenological bounds on model's parameters $u$ and $w$ by just looking at the static Coulomb 
interaction potential. 
\footnote{For $r\leq L < 1/M_Z$, one needs to take into account the full electroweak contribution. 
But if we neglected the contribution from the Z-boson and its KK modes as an illustrative computation, 
we would obtain to leading orders (\ref{leadingV}) 
$V(r)\simeq \frac{2{e^2}L}{{\pi}r^2}+ \frac{{\pi}e^2}{6L} -\frac{3 {e^2} L}{{\pi}wur^4} + \ldots $
It is interesting that, for $r\leq L$, the model-independent leading order of potential is $ \sim 1/r^2 $, 
which agrees with the classical result obtained by Gaussian theorem \cite{Hamed2} in a purely geometric 
approach.} 

\subsection{One-loop correction to KK mass}
It is known that in theories with just one extra dimension, the sum over an infinite tower of KK modes for 
tree-level diagrams converges as we have seen above. At the loop levels this is no longer true and 
certain renormalization procedure is needed. To see specifically how the KK spectrum obtained above 
and the KK-number non-conservation couplings contribute in an actual loop computation, let us evaluate, as 
an example, the fermion self-energy at one-loop and its corresponding mass shift in the 
4-d effective picture.

\begin{figure}
        \begin{center}
        \epsfig{figure=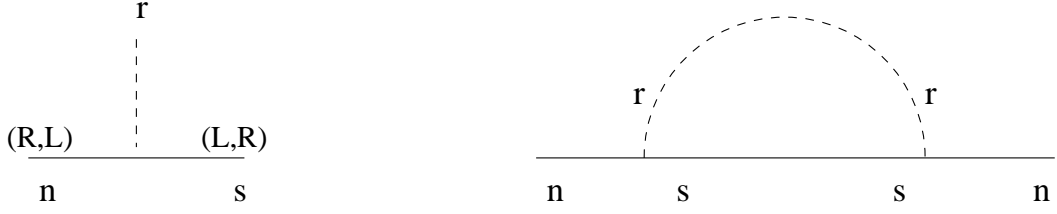,width=0.85\textwidth}
        \end{center}
        \vspace{0cm}
        \caption{Effective fermion-scalar vertex and fermion self-energy diagram}
        \label{Fig.5}
        \end{figure}

Decomposing $\psi (x,y)$ and $\phi (x,y)$ as in (\ref{chiraldecomp}), (\ref{scalarexpansion}) and 
performing integration over y-coordinate transform Lagrangian $ {\cal L}_{\psi} (\ref{LagPsi}) $ 
into its 4-d version
        \begin{displaymath}
\int{\cal L}_{\psi} dy= \sum_{n} { \bar{\psi} _{n} (x)  (\not {\partial} - m_{n}) \psi_{n} 
(x)} - \sum_{n,\bar{r},s} {f \bar{\psi}_{n} (x) \phi_{\bar{r}} (x) (g^{n\bar{r}s}_{RL} P_{L} + 
g^{n\bar{r}s}_{LR} P_{R}) \psi_{s} (x)}   
        \end{displaymath}   
where $m_{n}$'s are tree-level masses of KK fermions and $g$'s are effective 4-d couplings 
(Fig. \ref{Fig.5}).
        \begin{eqnarray}
        \label{g}
g^{n\bar{r}s}_{RL} &=& \int{\xi_{nR} (y) f_{\bar{r}} (y) \xi_{sL} dy}   \nonumber \\
g^{n\bar{r}s}_{LR} &=& \int{\xi_{nL} (y) f_{\bar{r}} (y) \xi_{sR} dy}
        \end{eqnarray}
We note that in $g^{n\bar{r}s}_{RL}$, the OBCs require $n$ and $s$ to be even and odd integer 
respectively. In $g^{n\bar{r}s}_{LR}$ however these parity constraints are reversed. The couplings 
are related by equation $g^{n\bar{r}s}_{RL}=g^{s\bar{r}n}_{LR}$.
The modified propagator of a fermionic KK mode $n$ is 
        \begin{displaymath}
\frac{1}{\not {p} - m_{n} - \Sigma_{n} (\not{p})}
        \end{displaymath} 
where
        \begin{eqnarray} 
        \label{sum}
-i\Sigma _{n} (\not p)&=&\sum_{\bar{r},s} {f^2 \int^{\Lambda} {\frac{d^4 k}{(2\pi)^4} 
\frac{1}{k^2-\bar{m}^2_r} (g^{n\bar{r}s}_{RL} P_{L} + g^{n\bar{r}s}_{LR} P_{R}) 
\frac{1}{(\not{p}-\not{k})-m_s} (g^{n\bar{r}s}_{LR} P_{L} + g^{n\bar{r}s}_{RL} P_{R}) }   }  \nonumber \\
&\rightarrow& \frac{i{f}^2}{16{\pi}^2} 
\sum_{\bar{r},s} { \left( \frac{1}{2} 
((g^{n\bar{r}s}_{RL})^2 P_R 
+ (g^{n\bar{r}s}_{LR})^2 P_L ) \not{p}   
+ g^{n\bar{r}s}_{RL} g^{n\bar{r}s}_{LR} m_{s} \right)}
\ln{\left( \frac{\Lambda^2}{m^2_{s}} \right)}
        \end{eqnarray}
where the last expression was obtained using Feynman's parameterization and Wick's rotation, 
$\Lambda$ is a cut-off scale, above which the physics is governed by a more fundamental theory, and the 
sum over KK modes is accordingly limited by the relations ${\bar{m}}^2_{r}, m^2_s \leq {\Lambda}^2$.

If ${\Lambda}^2 \geq w^2$, heavy KK modes (${\bar{m}}^2_{r} \geq u^2, m^2_s \geq w^2 $) 
will contribute to the sum (\ref{sum}) and drive its value to that of the UED scenario. In this work we 
assume ${\Lambda}^2 \leq w^2$ to investigate the contribution, if 
any, solely from distinctive lower part of KK spectrum of brane picture. This assumption is 
self-consistent because by tuning $u$ and $w$ we can push $\Lambda$ to sufficiently high 
scale, and non-renormalizable contribution is expected to be cut off by quantum gravitational effects
\cite{UED}. 
With this assumption, from orbifolding constraints (\ref{Mcondition1S}) and (\ref{Mcondition2S}), it 
follows that only lower KK scalar modes ($ \bar{r} \ll uL$) appear in the sum (\ref{sum}) as their 
masses are below the cut-off. The number of these relevant modes is $N_{s} \sim {\cal{O}}(1) $ in 
the limit $L \gg \frac{1}{u}$ considered here. This observation allows a rough evaluation of 
non-universal couplings $g^{n\bar{r}s}_{RL}$.  

\begin{figure}
        \begin{center}     
        \epsfig{figure=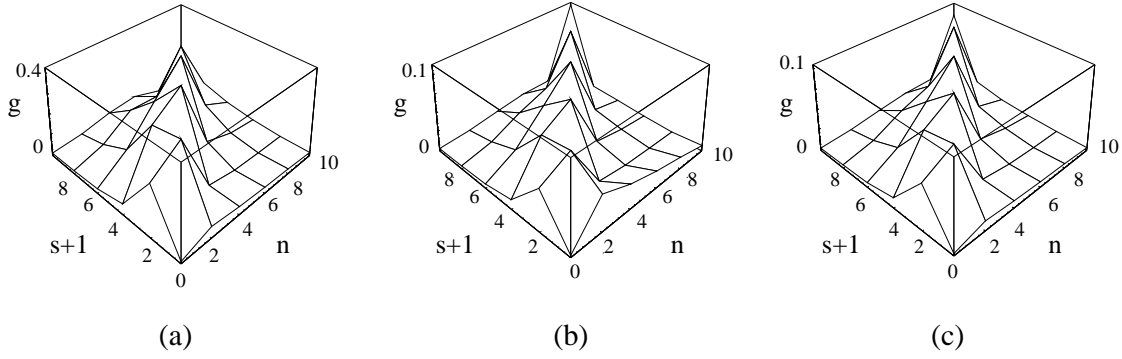,width=0.9\textwidth}
        \end{center}
        \vspace{0cm}
        \caption{Coupling $g^{n\bar{r}s}_{RL}$ (\ref{g}) as function of $n,s$ with $u=1$, $w/u=10.3$, 
         $uL=40$. The background scalar field $\phi _{\bar{r}}$ is in the normalizable (bound) KK 
         mode (Fig. 6a), or $\bar{r} = 0$ mode (Fig. 6b), or $\bar{r} = 5$ mode (Fig. 6c). The 
         position of the peaks clearly indicates $n=s+1$ enhancement. We note also that
         the last two graphs are very similar, 
         which justifies our approximation leading to (\ref{sigma}).}
        \label{Fig.6}
        \end{figure}

Specific solutions of $\xi_{nR,L}$ (\ref{ARBsol}), (\ref{ALBsol}) and $f_{\bar{r}}$ (\ref{Afsol}), 
(\ref{Afodd}) suggest that for KK modes relevant to the self-energy diagram ($n,s \leq \frac{w}{u}; 
\bar{r} \ll uL$), the characteristic width of normalized wavefunction of KK fermion $\xi_{nR,L}(y)$ in 
the extra dimension is much smaller than that of $f_{\bar{r}}(y)$ (of scalar field) as $w$ is 
sufficiently larger than $u$. 
\footnote{This assertion can be less 
rigorous for KK modes $s \rightarrow \frac{w}{u}$ in the upper limit of the sum, but contribution from 
these modes are actually suppressed by factor $\ln{\frac{\Lambda}{m_s}}$ } 
For a rough estimation we neglect the variation of $f_{\bar{r}}(y)$ over the extent of $\xi_{nR,L}(y)$, then 
for each relevant mode $\bar{r}$, couplings $g^{n\bar{r}s}_{RL}$ are most enhanced for few modes $s$ 
closest to $n$ (i.e. when $n \simeq s+1$), because 
it is when $\xi_{nR}$ resembles $\xi_{sL}$ most 
(i.e. $2\int_{0}^{L} \xi_{nR} \xi_{sL} dy \sim 1$). Along this 
approximation, the fermionic KK zero-mode mass should receives a rather small 1-loop correction, because 
it has no corresponding ``closest-neighbor'' mode $\xi_{sL}$ with $s=-1$, i.e. $g^{0\bar{r}-1}_{RL}
=g^{-1\bar{r}0}_{LR}=0$. Our estimation is verified by a 
particular numerical evaluation of coupling $g^{n\bar{r}s}_{RL}$ as function of mode's indices $n$, $s$. 
The result is presented in Fig. \ref{Fig.6}, where one can see clearly the effect 
of closest-neighbor enhancement on the value of couplings. The value of these couplings 
indeed decreases very quickly off the diagonal $n=s+1$. 
Putting all observations together, for KK non-zero mode of fermion we have 
        \begin{equation}
        \label{sigma} 
-i\Sigma _{n} (\not p) = {\cal{C}} \frac{i}{16\pi^2} \frac{f^2}{2L} 
\ln{\left( \frac{\Lambda^2}{m^2_{n}} \right)} \left (\frac{1}{2} \not{p}   +  m_{n} \right)
        \end{equation}
where ${\cal{C}} \sim {\cal{O}}(1)$ accounts for the number of relevant KK scalar modes in the sum and some 
crude estimation that we have made. We note also that $f^2/2L$ is just 4-d version of squared coupling 
$f^2$. This one-loop divergence requires the corresponding counter terms 
        \begin{displaymath}
{\cal L} \rightarrow {\cal L} + \delta {\cal L} = \sum_{n}
{\left( Z_{\psi} \bar{\psi}_{n} i\not{\partial} \psi_{n} - 
Z_{m} m_{n} \bar{\psi}_{n} \psi_{n} \right)} 
        \end{displaymath}
with renormalization scaling factors
        \begin{eqnarray}
Z_{\psi} &=& 1 - \frac{{\cal C}}{64 \pi^{2} }\frac{f^2}{2L}\ln{\left( \frac{\Lambda^2}{m^2_{n}} \right)}  
\nonumber \\
Z_{m}    &=& 1 + \frac{{\cal C}}{32 \pi^{2} }\frac{f^2}{2L}\ln{\left( \frac{\Lambda^2}{m^2_{n}} \right)}  
\nonumber
        \end{eqnarray}
After transforming to canonical basis $\psi \rightarrow \sqrt{Z_{\psi}} \psi$ we finally obtain an 
one-loop correction to the mass of fermionic KK non-zero mode (in the leading order)
        \begin{equation}
        \label{1-loop}
\delta m_{n} = m_{n} (\frac{Z_{\psi}}{Z_{m}}-1) 
\simeq m_{n} \left(  \frac{3{\cal C}}{64 \pi^{2} }\frac{f^2}{2L}\ln{ \frac{\Lambda^2}{m^2_{n}}} \right)
        \end{equation}  
where $m_{n}$'s are given by (\ref{mRquantized}). Interestingly, the same result is obtained 
in \cite{CMS}, where one-loop correction to KK mass in UED scenario was computed using fundamental 
5-d approach (see \cite{GGH2}). This would indicate that, at tree-level, KK mass structures in fat 
brane and UED scenarios are very different at least in the lower part as we have seen in previous 
sections, however their radiative correction may scale somewhat similarly. We particularly note that, 
OBCs and fat brane structure contribute crucially to the mass shift (\ref{1-loop}) at two places. 
First, the OBCs (\ref{Mcondition1S}), (\ref{Mcondition2S}) eliminate most of scalar KK states that 
otherwise would have appeared in the fermion self-energy diagram. Second, non-trivial interaction in 
extra dimension is manifested entirely in analytical properties of effective 4-d couplings 
$g^{n\bar{r}s}_{RL}$ that in turn enhance certain modes of mixing. This non-trivial mixing has 
certain important implications to Standard Model and that will be presented in the subsequent work. 
We now turn to some resonant effect on KK spectrum of fermions.   

\section{Resonances and Kaluza-Klein spectrum of fermions}
The consideration of section III has led to the conclusion that orbifold boundary conditions, which 
are crucial ingredients in achieving single-chirality Standard Model fermions as zero-modes of fields 
in a higher dimensional theory, act to suppress many bound ($m^2 < w^2$) fermionic KK levels. 
Each of these levels' wavefunctions has a definite parity with respect to a fixed point, and if this 
parity is not identical to the one imposed by OBCs, the corresponding level is suppressed by parity 
mismatch. Higher levels ($m^2 \geq w^2$) is two-fold degenerate and can be arranged to have the desired 
parities (Appendix A), so the impact of OBCs on these levels is not so obvious. In this section 
we discuss in more details the structure of this upper part of KK spectrum and clarify a 
close connection between periodic boundary conditions, orbifold boundary conditions and the complex 
transmission amplitude of free fermion through the double-brane system mentioned in section III.

We first consider a related problem of one non-compact dimension where a fermion, being free at infinity, 
approaches a single double-well potential $V_R$ (\ref{inftychiralV}) (Fig. \ref{Fig.2}) with wave vector 
$ku \equiv \sqrt{m^2-w^2} >0$. In this problem, the wavefunction of particle is not constrained by any 
boundary conditions since the dimension is non-compact. Because the double-well is not left-right 
symmetric, we expect that the complex reflection amplitudes are not the same for two opposite propagation 
directions. 
\footnote{However, complex transmission amplitudes through generic 1-d potential are identical for 
two propagation directions (see e.g. \cite{LL}). } 
The most general particle's wavefunction and its asymptotic forms as 
referred to the fixed point at $y=0$ are respectively (see (\ref{ARcont1}), (\ref{ARcont2}))
          \begin{eqnarray}
          \label{generalR}
\xi _{mR}(y)&=&aR_1(y)+bR_2(y)                  \\
          \label{assymgeneralR+}
y\rightarrow \infty  &:& \xi _{mR}(y)\rightarrow  a e^{ikuy} +  b e^{-ikuy}    \\
          \label{assymgeneralR-}
y\rightarrow -\infty &:& \xi _{mR}(y)\rightarrow (a C_1 + b C_2) e^{-ikuy} + (a D_1 + b D_2) e^{ikuy} 
          \end{eqnarray}
As referred to the fixed point at $y=L$, similarly we have
          \begin{eqnarray}
          \label{generalR'}
\xi '_{mR}(y)&=&a'L_1(y')+b'L_2(y')              \\
          \label{assymgeneralR'+}
y'\rightarrow \infty  &:& \xi '_{mR}(y)\rightarrow  a' e^{ikuy'} +  b' e^{-ikuy'}    \\
          \label{assymgeneralR'-}
y'\rightarrow -\infty &:& \xi '_{mR}(y)\rightarrow (a' C'_1 + b' C'_2) e^{-ikuy'} + 
(a' D'_1 + b' D'_2) e^{ikuy'} 
          \end{eqnarray} 
where $y'=y-L$ and $a, b, a', b'$ are constant coefficients. 

For the wave traveling from left to right and scattering on the double-brane, we let $b'=0$ and 
the matching of (\ref{assymgeneralR+}) and (\ref{assymgeneralR'-}) gives 
$\frac{a}{b}=\frac{D'_1}{C'_1}\exp{(-2ikuL)}$ and 
$\frac{a'}{b}=\frac{1}{C'_1}$. Next, using relations (\ref{ADrelations}) $C_1=-D_2$, $C_2=D ^* _1$ 
and $D'_2=-D_2=D ^* _2$ we obtain the transmission amplitude
          \begin{equation}
          \label{tR}
t_R=\frac{a'}{aD_1+bD_2}=\frac{1}{D_1D'_1\exp{(-2ikuL)}+D ^2 _2}     
          \end{equation} 
Now we see clearly that conditions (\ref{Mcondition1}), (\ref{Mcondition2}) imply the physical 
requirement that the complex transmission amplitude of a fermion propagating through a single 
double-well potential $V_R$ is precisely one. We will come back to this observation below. 
A similar consideration gives identical transmission amplitude for wavefunction subject to the 
other type of a single double-well potential $V_L$ (\ref{inftychiralV}): $t_L = t_R$. From here 
on we drop all indices $R,L$ as well as the factor $\exp{(-2ikuL)}$ in the expression of $t$ by virtue of 
(\ref{Mcondition1}).      
   
For $k \gg \frac{w}{u}>1$, we saw earlier that $m_n^2 \simeq \frac{n^2 \pi ^2}{L^2}$, i.e. the higher 
KK structure is always dominated by usual periodic compactification. Now we can see this result 
more physically: the high-lying KK modes have lagre transmission amplitude $t \rightarrow 1$ because they
are not sensitive to the underlying potential and can cross it without significant reflection, and 
on these modes, the periodic boundary conditions (\ref{Mcondition1}) are the more influential ones. 
However, as seen in previous section, the couplings $f$ and $\lambda$ in Eq. (\ref{Lag}) 
have negative dimensions of mass, so the theory is not renormalizable. It may effectively describe 
physics only under a certain mass scale, and our calculation may be no longer relevant for heavy KK 
modes above that scale. Apart from this, smaller range of values of $k$ deserves a special interest, 
also because it is where the fat brane structure is expected to play a dominant role. 
To see this specifically, we now examine the mass quantization equation $t=1$ (\ref{Mcondition2}) for 
low-lying fermionic KK states, $k \ll 1< \frac{w}{u}$. Using the product expansion \cite{HTF}
             \begin{displaymath}
\frac{\Gamma (z_1) \Gamma (z_2)}{\Gamma (z_1+z_3)\Gamma (z_2-z_3)}=
\prod_{q=0}^{\infty} (1+\frac{z_3}{z_1+q})(1-\frac{z_3}{z_2+q})
             \end{displaymath}
we can expand $D_1$ as follows
             \begin{eqnarray} 
             \label{expandD1}
D_1&=&\frac{k-i\frac{w}{u}}{k}\prod_{q=1}^{\infty}(1-\frac{w/u}{q-ik})(1+\frac{w/u}{q-ik})  \nonumber \\
   &=&|D_1|\exp {\left( -i\frac{\pi}{2} + i[\frac{w}{u}]\pi + ik\frac{u}{w}
+ ik\sum_{q=1}^{\infty}( \frac{2}{q}-\frac{1}{q-w/u} -\frac{1}{q+w/u} ) + {\cal O} (k^3) \right)} 
\nonumber \\
   &=&|D_1|\exp {\left( -i\frac{\pi}{2} + i[\frac{w}{u}]\pi + ik\frac{u}{w}
+ ik(2\gamma + \Psi(\frac{w}{u})+ \Psi(\frac{-w}{u})) + {\cal O} (k^3) \right)}  \nonumber
             \end{eqnarray}
where $[\frac{w}{u}]$ is the maximal integer not larger than $\frac{w}{u}$, 
$\Psi (z) \equiv \frac{d}{dz} \ln {\Gamma (z)} $ is PolyGamma function, and $\gamma \approx 0.577$ is
Euler-Mascheroni constant. After expanding $D'_1$ in a similar way and using recursion 
formula $\Psi (s+1)=\Psi (s) +\frac{1}{s}$ we obtain
             \begin{equation} 
             \label{expandD1D'1}
D_1D'_1=|D_1D'_1| \exp{(ikP(\frac{w}{u})+{\cal O}(k^3))}=|D ^2 _1| \exp{(ikP(\frac{w}{u})+{\cal O}(k^3))}
             \end{equation}  
with 
             \begin{equation} 
             \label{defofP}
P(\frac{w}{u}) \equiv 2 \left( 2\gamma + \Psi(\frac{w}{u})+ \Psi(\frac{-w}{u}) \right)   
             \end{equation}
Now putting this back into the expression (\ref{tR}) of transmission amplitude, we find
             \begin{equation} 
             \label{t}
t=\frac{1}{D_1D'_1+D ^2 _2}=\frac{1}{|D_1|^2\exp{(ikP(\frac{w}{u})+{\cal O}(k^3))}-|D_2|^2} 
             \end{equation}
where $|D_1|^2$, $|D_2|^2$ have been calculated in Appendix A
             \begin{equation}
             \label{squaredD1D'1D2}
|D_1|^2=\frac{\sinh ^2 {\pi k} + \sin ^2 {\pi \frac{w}{u}}}{\sinh^2 {\pi k}}  \; \; \; \; \; \;
|D_2|^2=\frac{ \sin ^2 {\pi \frac{w}{u}}}{\sinh^2 {\pi k}}
             \end{equation}             

\begin{figure}
        \begin{center}
        \epsfig{figure=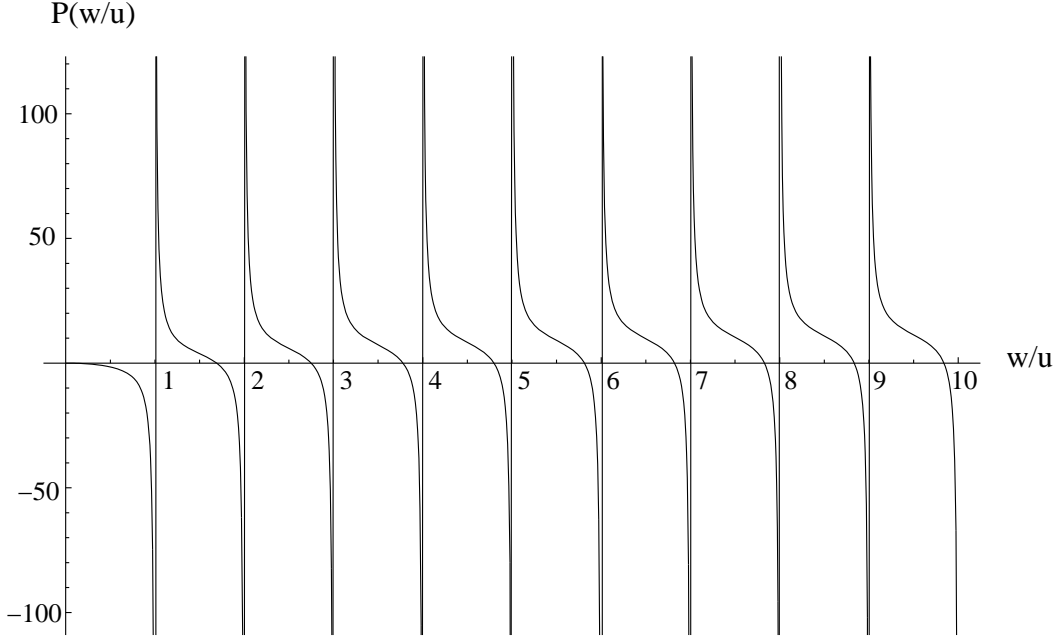,width=0.85\textwidth}
        \end{center}
        \vspace{0cm}
        \caption{Sketch of Function $P(w/u)\equiv 2 \left( 2\gamma + \Psi(w/u)+ \Psi(-w/u) \right) $}
        \label{Fig.7}
        \end{figure}

First, if $w/u$ has values such that $P(w/u)$ in (\ref{defofP}) or 
$(\sin ^2 {\pi \frac{w}{u}})$ in (\ref{squaredD1D'1D2}) vanishes, one can see that 
$t=1+{\cal O}(k)$, i.e. the condition 
(\ref{Mcondition2}) is approximately satisfied for low-lying KK modes 
$k \ll 1 $. Using condition (\ref{Mcondition1}), this means $n \ll uL$, and since we are considering 
the limit $uL \gg 1$, this approximation holds for a certain number of modes. This again can be seen 
as a resonance behavior of the potentials (\ref{inftychiralV}) that for certain values of $w/u$, 
even low-lying 
particles can essentially go through it, and condition (\ref{Mcondition1}) then determines their masses 
$m_n^2\approx w^2 + n^2{\pi}^2 / L^2$. The resonance values of $w/u$ are positive integers (for     
$\sin {(\pi \frac{w}{u})} = 0$) and others being solutions of equation $P(w/u)=0$. From Fig. \ref{Fig.7} 
we see that, there is exactly one such solution between any two successive integers. 
For all other values of $w/u$, we find that $t={\cal O}(k)$ and equation $t=1$ does not have solutions for 
$k \ll 1$. In other words, branes are almost ``opaque'' for low-lying modes and consequently, these 
fermion modes are absent in the KK spectrum. 

In overal, except for some particular values of bulk potential parameter $w/u$, we see that the lower 
part of KK spectrum of fermions is strongly distorted (i.e. suppressed) by orbifold compactification 
and the VEV of the background scalar field.  

\section{Conclusion}

In this work we attempt to investigate the effect of finite size of extra dimension, in the 
presence of a background scalar field, on masses and wavefunctions of KK modes of fermions and scalar fields 
by invoking the analog bulk potential, which is closely related to the brane picture generated by the 
kink solution of the scalar field equation. The size of these branes is finite (namely fat branes) 
and comparable to that 
of the kink. The compactification of the extra dimension to obtain a low-energy 
4-d realistic chiral theory involves different types of boundary conditions: 
the usual periodic boundary conditions which dominate the high-lying KK modes and give them the 
familiar structure of conventional compactification, and the orbifold boundary conditions, which 
along with non-trivial fat branes dominate the bound and low-lying KK modes, thus give them more a 
distinctive structure. The observation being emphasized here is that
these conditions on wavefunctions are not always compatible, depending on the specific values of the 
parameters of the model through some resonance effects. The general result is that, certain light 
fermionic KK modes are suppressed making incomplete the KK tower of fermions bound to each fixed points. 
In this respect, we see that scalar fields are less affected by OBCs, because they are 
not chiral. Roughly speaking, the limit separating the two very different parts of the spectrum 
is of the order of the 
potential barrier's height experienced by particles along the extra dimension, and this could 
serve as the cut-off scale required by a non-renormalizable higher dimensional theory. 
Integrating out the extra dimension leads to an effective 4-d theory. The effects of 
orbifold compactification and symmetry breaking in the transverse dimension are now embedded in the  
effective couplings, and the corresponding vertex interactions do not necessarily conserve the KK number. 
This in turn generates the contribution of KK modes to tree and higher-level processes.  
Making use of the analytical solutions of the KK masses and wavefunctions obtained in this paper, 
we will return to 
the implications of this brane picture to the Standard Model in a subsequent work. 
For an extra dimension of arbitrary size, certain numerical techniques are required to 
solve for the background scalar VEV subject to given orbifold boundary conditions. The matching of 
component wavefunctions is not simple, but we believe that the use of bulk potential would remain 
the right approach to find KK masses and their state functions in this general case.
\begin{acknowledgments}
PQH would like to thank Gino Isidori and the Theory Group at LNF (Frascati) for 
hospitality during the course of this work. N-KT would like to thank Dr. A. Soddu, Prof. V. Celli and 
Prof. P. Arnold for many helpful discussions and advice. This work is supported in part by the 
U.S. Department of Energy under Grant No. DE-A505-89ER40518.
\end{acknowledgments}
\begin{appendix}
\section{}
In this appendix, we solve and classify different solutions of differential Eqs. (\ref{QMXieqR}), 
(\ref{QMXieqL}), which have been
accordingly employed with different physical constraints in the main text.
First, let us introduce some short-hand notations, 
        \begin{displaymath}
w \equiv fv ; \;\;\;\;  u \equiv \sqrt{\frac{\lambda v^2}{2}} ; \;\;\;\; 
z \equiv \tanh { \sqrt{\frac{\lambda v^2}{2}} y}  = \tanh{uy} ; \;\;\;\; 
\varepsilon \equiv \sqrt{\frac{w^2-m^2}{u^2}}       
        \end{displaymath}
Dividing both sides of (\ref{QMXieqR}) by $u^2(1-z^2)$ we obtain
        \begin{equation}
        \label{A1zeq}
\frac{d}{dz} \left( (1-z^2) \frac{d \xi _{mR}}{dz} \right) + \left( \frac{w}{u}(\frac{w}{u}+1) - 
\varepsilon ^2 \frac{1}{1-z^2}\right) \xi _{mR} = 0
        \end{equation}
Again, using some new notations,
$ z_1 \equiv \frac{1}{2} (1-z)$, and $\xi _{mR}(z) \equiv (1-z^2)^{\varepsilon /2} p(z) $
we transform (\ref{A1zeq}) into the standard hypergeometric differential equation
        \begin{equation}
z_1(1-z_1)\frac{d^2 p}{d z_1 ^2} +(\varepsilon +1)(1-2z_1)\frac{d p}{d z_1} -
(\varepsilon - \frac{w}{u})(\varepsilon +  \frac{w}{u} +1)p = 0
        \label{Ahypereq}
        \end{equation}   
Eq. (\ref{Ahypereq}) has two linearly independent solutions \cite{HTF}
        \[
r_1(z_1)=F(\varepsilon-\frac{w}{u},\varepsilon+\frac{w}{u}+1,\varepsilon+1;z_1) \newline
r_2(z_1)=z_1 ^{-\varepsilon}F(-\frac{w}{u},\frac{w}{u}+1,-\varepsilon +1;z_1)
        \] 
where F is hypergeometric function,
        \begin{equation}
        \label{hyperfn}
F(a,b,c;z)=1+\frac{ab}{c} \, \frac{z}{1!}+\frac{a(a+1)b(b+1)}{c(c+1)} \, \frac{z^2}{2!}+...
        \end{equation}
Two corresponding solutions of Eq. (\ref{A1zeq}) are
        \begin{equation}
        \label{AR1}
R_1(y)=(1-z^2)^{\varepsilon /2} r_1(z) = 
(\frac{1}{\cosh{uy}})^\varepsilon F(\varepsilon - \frac{w}{u},\varepsilon + \frac{w}{u} + 1,\varepsilon 
+1;\frac{1}{1+e^{2uy}})
        \end{equation}  
        \begin{equation}
        \label{AR2}
R_2(y)=(1-z^2)^{\varepsilon /2} r_1(z) = 
(2e^{uy})^\varepsilon F(-\frac{w}{u},\frac{w}{u} + 1,-\varepsilon +1;\frac{1}{1+e^{2uy}})
        \end{equation}     
First, consider the discrete spectrum $m^2 < w^2$; $\varepsilon >0$. Asymptotic forms of (\ref{AR1}), 
(\ref{AR2}) are
        \begin{equation}
        \begin{array}{rl}
        \label{AassymR1+} 
y\rightarrow \infty  : &    R_1 (y) \rightarrow 2^\varepsilon e^{-\varepsilon uy} \\
        \label{AassymR1-}
y\rightarrow -\infty  : &   R_1 (y) \rightarrow 2^\varepsilon e^{\varepsilon uy} 
F(\varepsilon - \frac{w}{u},\varepsilon + \frac{w}{u} + 1,\varepsilon +1;1) 
        \end{array}
        \end{equation}
        \begin{equation}
        \begin{array}{rl}
        \label{AassymR2+} 
y\rightarrow \infty  : &    R_2 (y) \rightarrow 2^\varepsilon e^{\varepsilon uy} \\
        \label{AassymR2-}
y\rightarrow -\infty  : &   R_2 (y) \rightarrow 2^\varepsilon e^{\varepsilon uy} 
F(-\frac{w}{u},\frac{w}{u} + 1,-\varepsilon +1;1)
        \end{array}
        \end{equation}
When $L \rightarrow \infty$ as considered in section II, we see from (\ref{AassymR2+}) that $ R_2$ blows 
up at infinity and cannot be a physical solution for bound states , and 
        \begin{equation}
\xi _{mR}(y)=R_1 (y)=
(\frac{1}{\cosh{uy}})^\varepsilon F(\varepsilon - \frac{w}{u},\varepsilon + \frac{w}{u} + 1,\varepsilon +1;
\frac{1}{1+e^{2uy}})
        \label{ARBsol}
        \end{equation}
is the sought physical solution of $\xi _{mR}$ with the following condition (otherwise, $F$ in 
(\ref{AassymR1-}) and then $\xi _{mR}$ in (\ref{ARBsol}) blow up as $y\rightarrow -\infty$)  
        \begin{equation}
        \label{AnR}
\varepsilon - \frac{w}{u}=-n_R  \;\;\;\;\;\;\; (n_R \in N)
        \end{equation}
From this follows the mass quantization of KK discrete levels (\ref{mRquantized}). 
As far as the zero-mode ($m=0$) is concerned, in certain computation it is more convenient to 
approximate $\xi_{0R}$ to a normalized Gaussian function
        \begin{equation}
        \label{Gaussianfn}
\xi_{0R}(y)=(\frac{wu}{\pi})^{1/4} e^{-wuy^2/2}
        \end{equation}
The fact that here only $R_1$ is the physical solution is readily understood, as all bound states of 
a 1-d potential are non-degenerate. Further , we see from (\ref{ARBsol}) that 
$\xi _{mR}$ is even and odd function respectively when $n_R$ is even and odd integer. 
For the continuous spectrum, $m^2 \geq w^2$; $k \equiv \sqrt{\frac{m^2-w^2}{u^2}}=i \varepsilon$, 
(\ref{AR1}), (\ref{AR2}) read
        \begin{eqnarray}
        \label{ARcont1}
R_1(y)&=&(\frac{1}{\cosh {uy}})^{-ik}F(-ik-\frac{w}{u},-ik+\frac{w}{u}+1,-ik+1;\frac{1}{1+e^{2uy}}) \\
        \label{ARcont2}
R_2(y)&=&(2e^{uy})^{-ik} F(-\frac{w}{u},\frac{w}{u}+1,ik+1;\frac{1}{1+e^{2uy}})
        \end{eqnarray} 
whose asymptotic forms are
        \begin{equation} 
        \begin{array}{rl}
        \label{AassymR1cont+}
y\rightarrow \infty  : &  R_1 (y) \rightarrow 2^{-ik} e^{ikuy} \\
        \label{AassymR1cont-}
y\rightarrow -\infty  : &  R_1 (y) \rightarrow 2^{-ik}(C_1 e^{-ikuy} + D_1 e^{ikuy})
        \end{array}
        \end{equation}  
        \begin{equation}
        \begin{array}{rl}
        \label{AassymR2cont+}
y\rightarrow \infty  : &   R_2 (y) \rightarrow 2^{-ik} e^{-ikuy} \\
        \label{AassymR2cont-}
y\rightarrow -\infty  : &    R_2 (y) \rightarrow 2^{-ik}(C_2 e^{-ikuy} + D_2 e^{ikuy})
        \end{array}
        \end{equation}
where for the limit $y\rightarrow -\infty$, we have used a fundamental analytic continuation formula 
of hypergeometric function \cite{HTF}
        \begin{eqnarray}
F(a,b,c;z) &=& \frac{\Gamma (c) \Gamma (c-a-b)}{\Gamma (c-a) \Gamma (c-b)}F(a,b,a+b+1-c;1-z)  + \nonumber \\ 
           & & (1-z)^{c-a-b} \frac{\Gamma (c) \Gamma (a+b-c)}{\Gamma (a) \Gamma (b)}F(c-a,c-b,c+1-a-b;1-z)
        \end{eqnarray} 
and
        \begin{equation}
        \label{ACD}
        \begin{array}{lr}
C_1 = \frac{\Gamma (ik) \Gamma (1-ik)}{\Gamma (\frac{w}{u}+1) \Gamma (-\frac{w}{u})}      & 
D_1 = \frac{\Gamma (-ik) \Gamma (1-ik)}{\Gamma (-ik-\frac{w}{u}) \Gamma (-ik+\frac{w}{u}+1)}   \\
C_2 = \frac{\Gamma (ik) \Gamma (1+ik)}{\Gamma (ik-\frac{w}{u}) \Gamma (ik+\frac{w}{u}+1)}  &   
D_2 = \frac{\Gamma (-ik) \Gamma (1+ik)}{\Gamma (\frac{w}{u}+1) \Gamma (-\frac{w}{u})} 
        \end{array}
        \end{equation}

Using Gamma function's relations \cite{GR}; $\Gamma (z+1)=z \Gamma (z) $, 
$\Gamma (z) \Gamma(1-z)=\frac{\pi}{\sin{(\pi z)}}$, we obtain some relations useful for our calculation
         \begin{eqnarray}
         \label{ADrelations}
C_1=\frac{i \sin{ \pi \frac{w}{u}} }{\sinh{ \pi k}} & \Rightarrow &   
|C_1|^2=\frac{ \sin ^2 {\pi \frac{w}{u}} }{\sinh^2 {\pi k}}  \nonumber \\
C_1=-D_2 ;\;\;\;\;\;\;\;C_2=D ^* _1 & \Rightarrow & |C_1|^2=|D_2|^2 ; \;\;\;\;\;\;\; |C_2|^2=|D_1|^2 
\nonumber \\
|D_1|^2-|D_2|^2=|D_1|^2 + D^2 _2 =1 & \Rightarrow &
|D_1|^2=\frac{\sinh ^2 { \pi k} + \sin ^2 {\pi \frac{w}{u}} }{\sinh ^2 {\pi k}}
         \end{eqnarray}
We note specially that when $\frac{w}{u}$ is integer, $D_2=0$ and $|D_1|=1$.

In contrast with bound states, none of (\ref{ARcont1}), (\ref{ARcont2}) has definite parity. 
However, continuous levels of 1-d potential are always 2-fold 
degenerate, so we can linearly combine $ R_1$, $R_2$ to produce function of desired parity. From 
(\ref{AassymR1cont+})-(\ref{AassymR2cont-}) we can respectively build up the even and odd 
right-handed wavefunctions 
        \begin{eqnarray}
        \label{AReven}
R_{even}(y)=R_1 (y) + \frac{D_1}{1-D_2} R_2 (y)  \\
        \label{ARodd}
R_{odd}(y)=R_1 (y) - \frac{D_1}{1+D_2} R_2 (y)
        \end{eqnarray}
 
A similar consideration applies to the left-handed fermion wavefunctions, in place of (\ref{AR1}), 
(\ref{AR2}) we have
        \begin{eqnarray}
        \label{AL1}
L_1(y)&=&(\frac{1}{\cosh{uy}})^\varepsilon F(\varepsilon - \frac{w}{u}+1,\varepsilon + \frac{w}{u},
\varepsilon +1;\frac{1}{1+e^{2uy}})  \\
        \label{AL2}
L_2(y)&=&(2e^{uy})^\varepsilon F(-\frac{w}{u}+1,\frac{w}{u},-\varepsilon +1;\frac{1}{1+e^{2uy}})
        \end{eqnarray}
The key observation here, when comparing (\ref{AR1}), (\ref{AR2}) with (\ref{AL1}), (\ref{AL2}), 
is that left-handed solutions $L_1$, 
$L_2$ are effectively the same as $R_1$, $R_2$ after changing $\frac{w}{u}$ to $\frac{w}{u}-1$. 
Then the corresponding physical solutions and 
mass quantization equation for left-handed discrete spectrum are (see (\ref{ARBsol}), (\ref{AassymR1+}), 
(\ref{AnR}))           
        \begin{equation}
\xi _{mL}(y)=L_1 (y)=
(\frac{1}{\cosh{uy}})^\varepsilon F(\varepsilon - \frac{w}{u}+1,\varepsilon + \frac{w}{u},\varepsilon +1;
\frac{1}{1+e^{2uy}})
        \label{ALBsol}
        \end{equation}
        \begin{equation}
y\rightarrow \infty  : \;\;\;\;\;\;\;   L_1 (y) \rightarrow 2^\varepsilon e^{-\varepsilon uy} 
        \label{AassymL1}
        \end{equation}
        \begin{equation}
        \label{AnL}
\varepsilon - \frac{w}{u}+1=-n_L  \;\;\;\;\;\;\; (n_L \in N)
        \end{equation}
In the continuous spectrum, instead of (\ref{ARcont1})-(\ref{AassymR2cont-}), we have
        \begin{eqnarray}
        \label{ALcont1}
L_1(y)&=&(\frac{1}{\cosh {uy}})^{-ik}F(-ik-\frac{w}{u}+1,-ik+\frac{w}{u},-ik+1;\frac{1}{1+e^{2uy}}) \\
        \label{ALcont2}
L_2(y)&=&(2e^{uy})^{-ik} F(-\frac{w}{u}+1,\frac{w}{u},ik+1;\frac{1}{1+e^{2uy}})
        \end{eqnarray}
 
        \begin{equation}
        \begin{array}{rl}
        \label{AassymL1cont+} 
y\rightarrow \infty  : &  L_1 (y) \rightarrow 2^{-ik} e^{ikuy} \\
        \label{AassymL1cont-}
y\rightarrow -\infty  : &  
L_1 (y) \rightarrow 2^{-ik}(C'_1 e^{-ikuy} + D'_1 e^{ikuy})
        \end{array}
        \end{equation}

        \begin{equation}
        \begin{array}{rl}
        \label{AassymL2cont+}
y\rightarrow \infty  :   &  L_2 (y) \rightarrow 2^{-ik} e^{-ikuy} \\
        \label{AassymL2cont-}
y\rightarrow -\infty  : &  
L_2 (y) \rightarrow 2^{-ik}(C'_2 e^{-ikuy} + D'_2 e^{ikuy})
        \end{array}
        \end{equation}
where  
        \begin{equation}
        \begin{array}{lr}
C'_1 = \frac{\Gamma (ik) \Gamma (1-ik)}{\Gamma (\frac{w}{u}) \Gamma (-\frac{w}{u} +1)}      &
D'_1 =  \frac{\Gamma (-ik) \Gamma (1-ik)}{\Gamma (-ik-\frac{w}{u}+1) \Gamma (-ik+\frac{w}{u})} \nonumber  \\
C'_2  = \frac{\Gamma (ik) \Gamma (1+ik)}{\Gamma (ik-\frac{w}{u}+1) \Gamma (ik+\frac{w}{u})}  &   
D'_2 =  \frac{\Gamma (-ik) \Gamma (1+ik)}{\Gamma (\frac{w}{u}) \Gamma (-\frac{w}{u}+1)}
        \end{array} 
        \label{AC'D'}
        \end{equation}
By making the change $\frac{w}{u} \rightarrow \frac{w}{u} -1 $ in (\ref{ADrelations}) we can find 
similar properties of $C'_1$, $C'_2$, $D'_1$, $D'_2$. Specially, we have $D'_2=-D_2$, $|D'_1|=|D_1|$.  
The even and odd left-handed wavefunctions now are
        \begin{equation}
        \label{ALeven}
L_{even}(y)=L_1 (y) + \frac{D'_1}{1-D'_2} L_2 (y)
        \end{equation} 
        \begin{equation}
        \label{ALodd}
L_{odd}(y)=L_1 (y) - \frac{D'_1}{1+D'_2} L_2 (y)
        \end{equation}
Finally, for scalar field, Eq. (\ref{QMfeq}) can be solved by the same method. For discrete spectrum, 
$\bar{m} ^2 < 4u^2$, 
$\bar{\varepsilon} \equiv \sqrt{\frac{4u^2-\bar m ^2}{u^2}}>0$, we obtain respectively scalar 
wavefunction and mass quantization equation 
        \begin{equation}
f_{\bar n}(y)=(\frac{1}{\cosh{uy}})^{\bar{\varepsilon}} F(\bar{\varepsilon} - 2,\bar{\varepsilon} + 3,
\bar{\varepsilon} +1;\frac{1}{1+e^{2uy}})
        \label{Afsol}
        \end{equation}
        \begin{equation}
        \label{Anf}
\bar{\varepsilon} - 2=-\bar n \; \; \; \; \; \; \; (\bar n \in N)
        \end{equation}
Since $\bar{\varepsilon} >0$, there are only 2 discrete levels $\bar n =0 $, $1$ and the corresponding 
states are symmetric and antisymmetric at $y=0$.
In the continuous spectrum, 
$\bar m ^2 \geq 4u^2$, $\bar k \equiv \sqrt{\frac{\bar m ^2 - 4u^2}{u^2}}=i\bar{\varepsilon}$, in place 
of (\ref{ARcont1})-(\ref{AassymR2cont-}), (\ref{AReven}), (\ref{ARodd}), we have
        \begin{eqnarray}
S_1(y)&=&(\frac{1}{\cosh{uy}})^{-i \bar k}F(-i \bar k - 2,-i \bar k + 3,-i\bar k +1;\frac{1}{1+e^{2uy}}) \\
S_2(y)&=&(2e^{uy})^{-i\bar k} F(-2,3,i\bar k +1;\frac{1}{1+e^{2uy}})
        \label{Afcont}
        \end{eqnarray}
        \begin{equation} 
        \label{Aassymf1cont}
        \begin{array}{rl}
y\rightarrow \infty  : &  S_1 (y) \rightarrow 2^{-i\bar k} e^{i\bar k uy} \\
y\rightarrow -\infty : &  S_1 (y) \rightarrow 2^{-i\bar k}(\bar D e^{i\bar k uy})
        \end{array}
        \end{equation}
        \begin{equation}
        \label{Aassymf2cont}
        \begin{array}{rl}
y\rightarrow \infty: & S_2 (y) \rightarrow 2^{-i\bar k} e^{-i\bar k uy} \\
y\rightarrow -\infty: &  S_2 (y) \rightarrow 2^{-i\bar k}(\frac{1}{\bar D} e^{-ikuy})
        \end{array}
        \end{equation}
        \begin{equation}
        \label{Afeven}
S_{even}(y)=S_1 (y) + \bar D S_2 (y)
        \end{equation} 
        \begin{equation}
        \label{Afodd}
S_{odd}(y)=S_1 (y) - \bar D S_2 (y)
        \end{equation}
where 
        \begin{equation}
\bar D = \frac{\Gamma (-i\bar k) \Gamma (1-i\bar k)}{\Gamma (-i\bar k -2) \Gamma (-i\bar k +3)}=
\frac{(1+i\bar k)(2+i\bar k)}{(1-i\bar k)(2-i\bar k)} 
        \label{ADbar}
        \end{equation}
\end{appendix}

\end{document}